\documentstyle[aps,multicol,epsf]{revtex}

\begin{document}
\draft
\title{How sandpiles spill: Sandpile problem in a thick flow regime}

\author{S.N. Dorogovtsev$^{1, 2, \ast}$ and J.F.F. Mendes$^{1,\dagger}$}

\address{
$^{1}$ Departamento de F\'\i sica and Centro de F\'\i sica do Porto, Faculdade de Ci\^encias, 
Universidade do Porto\\
Rua do Campo Alegre 687, 4169-007 Porto -- Portugal\\
$^{2}$ A.F. Ioffe Physico-Technical Institute, 194021 St. Petersburg, Russia 
}

\maketitle

\begin{abstract}
We obtain an analytical solution of a one-dimensional sandpile problem in a thick flow regime, when it can be formulated in terms of linear equations. It is shown that a space periodicity takes place during the sandpile evolution even for a sandpile of only one type of particles. Similar periodicity was observed previously for many component sandpiles.  Space periods are proportional to an input flow of particles $r_0$. We find that the surface angle $\theta$ of the pile reaches it's final critical value ($\theta_f$) from lower values only at long times. The deviation ($\theta_f  -  \theta $) behaves as $(t/r_{0})^{-1/2}$.
\end{abstract}


\begin{multicols}{2}

\narrowtext

The increasing interest of the physics community in the study of granular materials is in one part related with the technological applications of such materials, but also, because of its unusual properties \cite{beh,kadan,ed1,her1,g1,duran,jaeger,frette}.
Recently, a new intriguing phenomena -- spontaneous stratification of granular mixtures while they are poured in a quasi-two-dimensional Hele-Shaw cell -- was observed \cite{makse97n,makse97prl,makse97pre,makse981,ciz,makse983,karol}. After the pouring, grains appear to be arranged in periodical layers of different species parallel to the pile surface. Several studies were presented in order to explain the 
stratification phenomenon; however a simpler question may be asked: is it possible to observe something similar to this one in the much simpler case of only one type of grains? The answer to this question  is not a easy task, since the sandpile problem is complicated even for such a situation. Fortunately, recent studies of granular flows made significant progress from the theoretical point of view \cite{bouch}. More recently, Boutreux, Rapha\"{e}l and P.-G. de Gennes \cite{bout1} proposed a phenomenological description of some special case of granular flows  -- a so called thick flow regime -- that admits an analytical treatment. In papers\cite{bout1,bout2} this approach was utilized to consider some granular flow configurations. 

We shall use the proposed phenomenological equations to answer positively the stated above question and to describe a total evolution of the sandpile, that will turn to be surprisingly non trivial. We will show that the slope of the pile approaches its critical value only at long times after a rather complicate relaxation process. In fact, we shall reconsider a classical sandpile problem using, may be, the simplest possible idea without appealing to more refined approaches like, for instance, self-organized criticality \cite{bak}.

The authors of paper \cite{bout1} proposed to describe phenomenologically one-dimensional thick granular flows (the flow thickness supposed to be much higher than the grain size) by the following equations:

\begin{eqnarray}
\label{e1}
\frac{\partial r}{\partial t} - v \; \frac{\partial r}{\partial x}   & = &  v_{u} \left( \frac{\partial h}{\partial x} - \theta_{f} \right) , \nonumber \\
\frac{\partial h}{\partial t} &  =  & - v_{u} \left( \frac{\partial h}{\partial x} - \theta_{f} \right) .
\end{eqnarray}
Here $h(x,t)$ is the profile of the static part of the material; $r(x,t)$ is the width of a moving granular layer; we assume that the flow is from right to left, all rolling grains are supposed to move with an equal velocity $v$. 
The sum of Eqs. (\ref{e1}), $\partial_{t}(r+h)-v\partial_{x} r = 0$, has the form of the continuity equation for grains. The right hand parts describe the conversion of the static grains to rolling ones and vice versa, the meaning of $v_u$ is the velocity of the uphill fronts, as we shall see later. Usually $v_{u} > v$ \cite{ciz}. $\theta_f$ is a critical angle, or a so called repose angle, that is the angle to which the sandpile will evolve. 
The physical reasons to introduce equations (\ref{e1}) in such a form are described in \cite{bout1,bout2}. 
These linear equations are much simpler for an analytical treatment than the previously proposed nonlinear equations for a thin flow regime \cite{bouch,bout01,bout02,bout03}. The general solution of equations (\ref{e1}) may be written immediately in the form \cite{bout1}:
\begin{eqnarray}
\label{e2}
r(x,t)  & = & u(x+vt) - \frac{v_u}{v + v_u} w(x - v_u t) - v_u \theta_f t,   \nonumber \\
h(x,t)  & = & w(x-v_u t) + v_u \theta_f t, \;\;\;\;\; r > 0;  \\[7pt]
h(x,t)  & = & \mbox{const}, \;\;\;\;\;\;\;\;\;\;\;\;\;\;\;\;\;\;\;\;\;\;\;\;\;\;\;    r = 0, \nonumber
\end{eqnarray}
where $u(x)$ and $w(x)$ are arbitrary functions.
We shall use it to describe phenomenologically the sandpile evolution neglecting, as usual \cite{bout1,bout2}, possible near front deviations from the thick flow regime.
Let there be a wall at $x=0$, and grains be pouring permanently from the moment $t=0$ at this point, so $r(x=0, t=0)=r_0$ is a boundary condition ($r_0$ is the thickness of the income flow). The sandpile is supposed to expand to the left, i.e. to $x = -\infty$. There are no particles  at first moment so $r(x\le 0, t=0) =0$, $h(x\le 0, t=0) =0$ are the initial conditions. Using Eqs. (\ref{e2}) and the above initial and boundary conditions one may obtain the following solution of Eq. (\ref{e1}) with a front moving to the left with the velocity $v$ at the time interval $0<t<t_{1} = (v+v_u)r_0/(v v_u \theta_f)$ [see also Fig.1(a)]: 
\begin{figure}
\epsfxsize=75mm
\epsffile{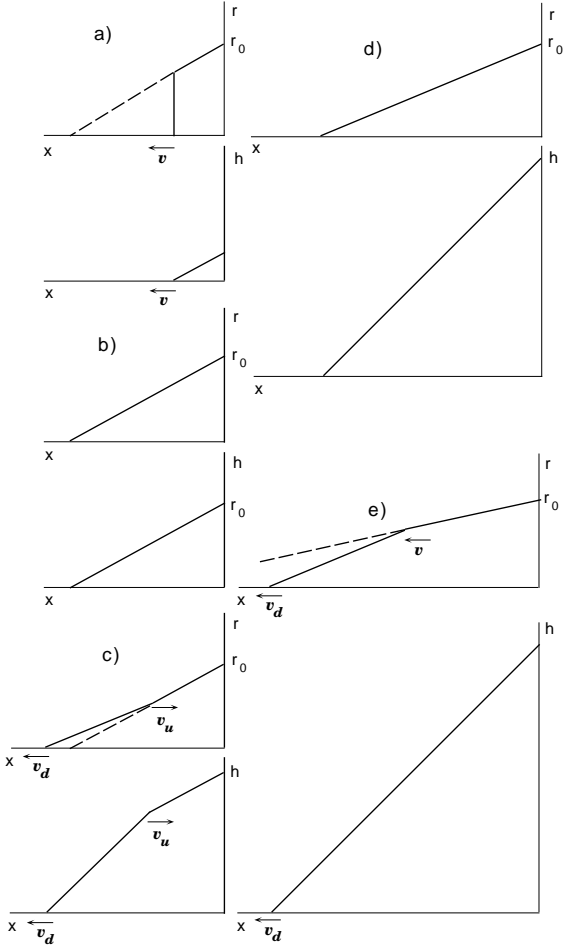}
\caption{
The evolution of the profiles of rolling grains $r(x)$ and static grains $h(x)$ at the region $-\infty<x<0$.  $r(-\infty<x<0,t=0)=h(-\infty<x<0,t=0)=0$. 
({\it a})  $0<t<t_1 $, the front moves with the velocity $v$.
({\it b})  $t=t_1 $ [see Fig. 2 and Eq. (4)].
({\it c})  $t_1<t<t_2 $, the front moves with some velocity $v_d<v<v_u$, the breaks of the profiles move uphill with the velocity $v_u$.
({\it d})  $t=t_2$, the breaks approach the wall at $x=0$.
({\it e})  $t_2<t<t_3$, the front proceeds to move with the velocity $v_d$, the break of $r(x)$ moves downhill with the velocity $v$. Note, that the right linear part of $r(x)$ is always motion less. After the break of $r(x)$ overtakes the front at $t=t_3$, the general configurations ({\it b}) -- ({\it e}) will repeat with a lower front velocity.
}
\end{figure}

\begin{eqnarray}
\label{e3}
r(x,t) & = & \left(r_0 + \frac{v_u}{v+v_u} \; \theta_{f} x \right) \Theta(x+vt) 
\nonumber \\[3pt]
h(x,t) & = & \frac{v_u}{v+v_u} \; \theta_{f}(x+vt)
\end{eqnarray}
$\Theta(x)$ is the Heaviside function (we do not write the multiplier $\Theta(x+vt)$ in the right hand side part of the second equation of Eqs. (\ref{e3}), since, of course, $r(x)$ and $h(x)$ can not be negative). Note, that the slope of the static part $\partial h/\partial x = v_u \theta_f/(v+v_u)$ is less than the critical slope.

As a result, at the time $t_1$ one has,

\begin{equation}
\label{e4}
r(x,t) = h(x,t) = r_0 + \frac{v_u}{v+v_u} \; \theta_{f} x
\end{equation}
for $-[(v+v_u )/(vv_u )]r_0/\theta_f<x<0$ (Fig. 1(b)). These equations are used as an initial condition to find the solutions for the time interval $t_1 < t < t_3$ (the time $t_3$ will appear naturally 
from the solution, and $t_2$ is an intermediate time, the meaning of it can be understood from Fig. 1). We have to add the following boundary conditions:

\begin{eqnarray}
\label{e5}
r(x=0,t>0) & = & r_0 ,  \nonumber \\
r\left(x=-\frac{v+v_u }{vv_u }\frac{r_0}{\theta_f}-v_d(t-t_1),t_1<t<t_3\right)  & = & 0 \nonumber \\
h\left(x=-\frac{v+v_u }{vv_u }\frac{r_0}{\theta_f}-v_d(t-t_1),t_1<t<t_3\right)  & = & 0 .
\end{eqnarray}
The second and the third equations of Eqs. (\ref{e5}) are the conditions for the left front of the pile that is supposed to move with an (unknown yet) velocity $v_d$. 

From the condition (\ref{e4}) and (\ref{e5}) and Eq.(\ref{e2}) we find the value $v_d = v_u/(v+2v_u)$ for $t_1 <t <t_3 $ and restore the evolution of the pile in this time interval [see Fig.1(b-e)]. Note that the profile breaks move uphill with velocity $v_u$ and move downhill with velocity $v$. Then we repeat the described procedure for the next time interval $t_3 < t < t_5 $, etc.

After these simple but rather tedious calculations we obtain a total solution consisting of linear parts, a structure of which one can see from  Figs. 1 and 2 and following equations. Lowest segmented line in Fig. 2 shows the dependence on time of the front position. Coordinates of the segments are:
\begin{eqnarray}
\label{e6}
x(t_{2m-1}<t<t_{2m+1}) = -\frac{m(m+1)}{2} \frac{(v+v_u)^{2}}{v_{u} {\cal V}_{m}} \frac{r_0}{\theta_f} \nonumber \\[3pt]
- \frac{v v_u}{{\cal V}_{m}} t, \;\;\; m=0,1,2,\dots
\end{eqnarray}
(for $m=0$ the time interval is $0<t<t_{1}$) with ${\cal V}_{m} \equiv mv + (m+1)v_u $. The particular times shown in Fig. 2 are

\begin{eqnarray}
\label{e7}
t_{2m-1} & = & \frac{m}{2} \frac{(v+v_u)[(m-1)v+(m+1)v_{u}]}{v v_{u}^2} \frac{r_0}{\theta_f} , \nonumber \\[3pt]
t_{2m}   & = & \frac{m(m+1)}{2} \frac{(v+v_u)^2}{v v_{u}^2} \frac{r_0}{\theta_f} ,  \\[5pt]
m        & = &1,2,\dots   \nonumber
\end{eqnarray}
The front coordinates corresponding to times $t_{2m-1}$ at which the front velocity changes its value are

\begin{equation}
\label{e8}
x(t_{2m-1}) = -m \frac{(v+v_u)}{v_{u}} \frac{r_0}{\theta_f} \ , \ m=1,2\ldots
\end{equation}
Note that they are arranged periodically. Two other types of lines are shown in Fig. 2: solid lines $x=v_u (t-t_{2m})$ and dashed lines $x=-v(t-t_{2m}), m=1,2,\ldots$ The lines of the first type depict the uphill movement of the breaks of both profiles $r(x)$ and $h(x)$ 
[Fig.1(c)]. The lines of the second type show the downhill movement of 
the break of the profile $r(x)$ with the velocity $v$ -- there is no any break of $h(x)$ in this situation [see Fig.1(e)]. Thus, in the problem under consideration, there exist both breaks of $r(x)$ and $h(x)$ moving uphill with the velocity $v_u$ and only the break of $r(x)$ moving downhill with the velocity $v$.

Solutions for all regions of Fig. 2, which are connected at these lines, look like:
\end{multicols}
\widetext
\noindent\rule{20.5pc}{0.1mm}\rule{0.1mm}{1.5mm}\hfill
\begin{eqnarray}
\label{e9}
r\left( -v(t-t_{2m-2}),v_u (t-t_{2m})<x<0 \right)  =  r_0 + \frac{v_u}{m(v+v_u)} \; \theta_f x, \nonumber \\
r\left(-\frac{m(m+1)}{2}\frac{(v+v_u)^2}{v_u {\cal V}_{m}} \frac{r_0}{\theta_f}-\frac{v v_u} {{\cal V}_{m}} t<x< -v(t-t_{2m}),v_u (t-t_{2m}) \right)  =  \frac{r_0}{2}+ \frac{1}{m(m+1)}\frac{v_u {\cal V}_{m}}{(v+v_u)^2} \; \theta_f \left(x+\frac{vv_u} {{\cal V}_{m}} t \right), \\[3pt]
m         = 1,2,\dots   \nonumber 
\end{eqnarray}
\begin{equation}
\label{e10}
h(v_u(t-t_{2m+2})<x<v_u (t-t_{2m}),0)  = \frac{m}{2}\frac{v+v_u}{v_u}r_0 + \frac{{\cal V}_{m}} {(m+1)(v+v_u)} \; \theta_f \left(x+\frac{vv_u}{{\cal V}_{m}}t \right), \;\;\; m = 0,1,2,\dots  
\end{equation}
\hfill\rule[-1.5mm]{0.1mm}{1.5mm}\rule{20.5pc}{0.1mm}
\begin{multicols}{2}
\narrowtext
Here, $t_0=0$. Inequalities in the right hand parts of Eqs. (\ref{e9}) and (\ref{e10}) define the areas of validity of the solutions [see Fig. 2].
Eqs. (\ref{e9}) and (\ref{e10}) describe totally the sandpile evolution, see Fig. 1(b-e).

\begin{figure}
\epsfxsize=70mm
\epsffile{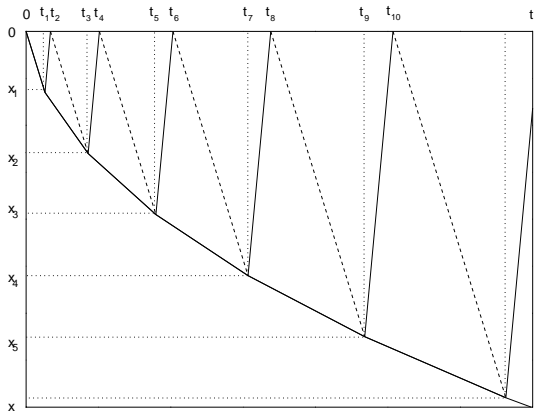}
\caption{
The areas of different solutions for the profiles of rolling and static grains [see Eqs. (\protect\ref{e9}) and (\protect\ref{e10})]. $v/v_u=0.3$\,. The lower segmented line shows the dependence of the front coordinate on time. The solid lines $x=v(t-t_{2m}), m=1,2,\ldots$ depict the uphill movement of the $r(x)$ and $h(x)$ breaks. The dashed lines $x=-v_u(t-t_{2m})$ show the downhill movement of the break of the $r(x)$ profile. The points $x_m$ are arranged periodically. 
}
\end{figure}

From Eq. (\ref{e10}) and Fig. 2, a space periodicity of the process is evident: general shapes of the profiles $r(x)$ and $h(x)$ are repeated each time when the front moves by the distance $(v+v_u)r_0/(v_u \theta_f)$ to the left. In fact, the pile is increased layer by layer, and  the expression for the width of these layers is the same as the one for the width of stratified layers of different fractions in a two-component sandpile that was obtained in papers \cite{makse97prl,ciz}. One may also see from Eq. (\ref{e10}) that heights of the pile at times $t_{2m}$ are also periodic in $m$:

\begin{equation}
\label{e11}
h(x=0,t_{2m}) = m \; \frac{v+v_u}{v_u} \; r_0 \ , m=0,1,2\ldots
\end{equation}

It follows from Eq. (\ref{e10}) that the slope, $\theta \equiv \partial h / \partial x$, of the pile will approach its critical value only at infinite time:

\begin{equation}
\label{e12}
\theta(t_{2m-3}<t<t_{2m}) = \left(1- \frac{v}{m\,(v+v_u)}\right)\theta_f \ , m=1,2\ldots
\end{equation}
(for $m=1, \ 0<t<t_2$) [see Fig. 3]. At $t_{2m-1}<t<t_{2m}, m=1,2,\ldots$ there are two different slopes for two parts of the profile $h(x)$. For $t_{2m}<t<t_{2m+1},m=0,1,2,\ldots$ all the profile has the same slope. Thus, for long times $t\gg[(v+v_u)^2/(vv_u^2)]r_0/\theta_f$ the slope behaves as

\begin{equation}
\label{e13}
\theta \cong \left(1-\frac{v}{v_u} \sqrt{\frac{r_0}{2v\,\theta_f t}\,}\right)\theta_f 
\ , 
\end{equation}
and $\theta$ relaxes slowly to its final value $\theta_f$ by a power law. At long times, the coordinate of front is obviously $x\cong\sqrt{2 v r_0 t/\theta_f}$, and its velocity tends to $-\sqrt{v r_0/(2\theta_f t)}$.

We presented the solution to the case of zero initial conditions -- there were no grains at the first moment. The case when there is already a pile with the critical slope at $t=0$ was also considered by us \cite{dm}. One may show that, after some period during which there still exists a part of the pile surface with the critical slope, we again will come to the considered above situation. Thus, at long times one may observe the same behavior for both types of initial conditions.

In summary, we demonstrated in the case of the thick flow regime that a space periodicity takes place during a sandpile evolution even in the case of the one-component pile. The pile spills during a repeating process: grains are piling layer by layer. The thickness of the layers coincides with the thickness of stratified layers at the two-component sandpile problem \cite{makse97prl,ciz}. Thus, in the one-component pile, we found a precursor of the stratification phenomena. 

\begin{figure}
\epsfxsize=50mm
\epsffile{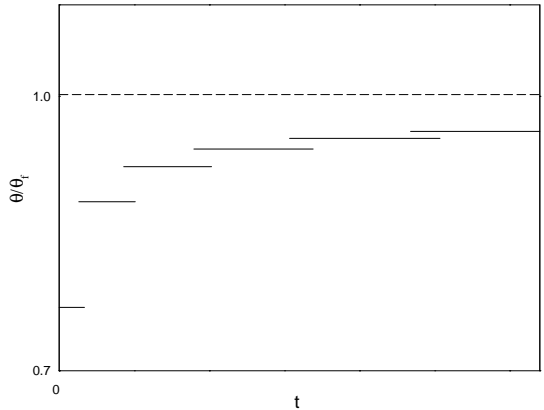}
\caption{
The dependence of the relative pile slope on time. $v/v_u=0.3$\,. At infinity $\theta \to \theta_f$. The lines are defined for $0<t<t_2$, $t_1<t<t_4$, $t_3<t<t_6$, etc. At times $t_1<t<t_2$, $t_3<t<t_4$, etc., the profile of static grain distribution has two parts with different slopes.}
\end{figure}

The slope of the pile goes to its final critical value after a complicated relaxation. At long times $\theta_f-\theta \propto \sqrt{r_0/t}$. Formally speaking, we studied only the one-dimensional problem, but an ordinary three-dimensional sandpile with an axial symmetry (that means that sand is pouring on an infinite horizontal plane at a single point) may be described by the same equations as Eqs. (1) (the coordinate $x$ means the distance from the center now). Thus, our results stay also valid for such a pile. The following questions remain open. How crucially do our results depend on the used phenomenological approach? Do the sandpile evolution in a thin flow regime differ essentially from the described one? \\

SND thanks PRAXIS XXI (Portugal) for a research grant PRAXIS XXI/BCC/16418/98. JFFM was partially supported by the projects PRAXIS/2/2.1/FIS/299/94, PRAXIS/2/2.1/FIS/302/94 and NATO grant No. CRG-970332. We also thank M.A. Santos for reading the manuscript and A.V. Goltsev  and A.N. Samukhin for many useful discussions.\\
$^{\ast}$      Electronic address: sdorogov@fc.up.pt\\
$^{\dagger}$   Electronic address: jfmendes@fc.up.pt

\end{multicols}
\end{document}